\documentclass[twocolumn]{aa}
\usepackage{graphicx,epsf,psfig,epsfig}
\begin{document}
   \title{IBIS performances during the Galactic Plane Scan \thanks{Based on observations with INTEGRAL, an ESA project
with instruments and science data centre funded by ESA member states (especially the PI countries:
Denmark, France, Germany, Italy, Switzerland, Spain), Czech Republic and Poland, and with participation of Russia and USA.}}

   \subtitle{I. The Cygnus region}

   \author{M. Del Santo\inst{1}, J. Rodriguez\inst{2,3}, P. Ubertini\inst{1}, A. Bazzano\inst{1}, A.J. Bird\inst{4}, 
F. Capitanio\inst{1}, L. Foschini\inst{5}, A. Goldwurm\inst{6}, F. Lebrun\inst{6},
A. Paizis\inst{3,7}, A. Segreto\inst{8}
}

\offprints{Melania Del Santo \\ delsanto@rm.iasf.cnr.it}

\institute{ IASF-CNR, via del Fosso del Cavaliere 100, 00133 Roma, Italy
\and  CEA Saclay, DSM/DAPNIA/SAp (CNRS FRE 2591), F-91191 Gif sur Yvette Cedex, France
\and  Integral Science Data Center, Chemin d'Ecogia 16, CH-1290 Versoix, Switzerland
\and  School of Physics and Astronomy, University of Southampton, Highfield, Southampton, SO17 1BJ, UK
\and  IASF/CNR, sezione di Bologna, via P. Gobetti 101, 40129 Bologna, Italy        
\and  CEA Saclay, DSM/DAPNIA/SAp, F-91191 Gif sur Yvette Cedex, France
\and  IASF/CNR, sezione di Milano, via Bassini 15, I-20133 Milano, Italy
\and  IASF/CNR, sezione di Palermo, via Ugo La Malfa 153, 90146 Palermo, Italy
        } 

\authorrunning{M. Del Santo et al.}
\titlerunning{Galactic Plane Scan -- Cygnus Region}

   \date{Received; accepted}

   \abstract{The Plane of our Galaxy is regularly observed by the INTEGRAL satellite by means of scheduled scans. 
We present here results from the IBIS/ISGRI instrument using data from the first two Galactic Plane Scans
performed at the end of the Performance Verification phase, while INTEGRAL was pointed towards the Cygnus region.

Considering the predicted IBIS sensitivity, we expected three persistent sources to be detectable: Cyg X--1, Cyg X--3, Cyg X--2 ,
in order of decreasing intensity in the hard-X energy range ($>$15 {\rm keV}).
In addition to these sources, our analysis revealed two more transient sources, 
confirming the unprecedented IBIS sensitivity. 
For each exposure ($\sim$2200 {\rm s}) of the two scans, we report on detected source fluxes, variabilities
and localisation accuracies.

   }

   \maketitle

\section{Introduction}
IBIS (\cite{ube03}) is the coded mask telescope on board of the INTEGRAL satellite (\cite{win03_1}) dedicated to
fine imaging ($12'$ angular resolution FWHM) of gamma-ray sources.
Thanks to its large field of view (29$^\circ \times$29$^\circ$ zero response) and wide energy range 
(15 {\rm keV}--10 {\rm MeV}), IBIS will monitor the gamma-ray sky for the next 2+3 years, depending on ESA approvals for
extension of the operations.
During the first year, the INTEGRAL observing programme is dedicated for $65\%$ of the time
to the General Programme (open time) and for $35\%$ to the Core Programme (CP), that is the guaranteed time reserved to the INTEGRAL
Science Working Team (ISWT).
    
The CP consists of three parts:
\begin{itemize}
\item[-]GCDE, Galactic Centre Deep Exposure
\item[-]GPS, Galactic Plane Scan
\item[-]Pointed observations, including Targets of Opportunity (ToOs) 
\end{itemize} 
In the GPS, the Galactic plane is scanned performing a ``$zigzag$'' pattern of $\sim$37 {\rm min} individual exposures 
separated by $6^\circ$ with respect to the scan path (\cite{win03_2}). 
Additionally, in the GCDE the Galactic Centre region will be observed deeply by closer 
(separations $2.4^\circ$ and $1.2^\circ$ in latitude and longitude, respectively) 
and shorter ($\sim$30 {\rm min}) pointings.

During the GPS programme the main IBIS goals are: discovering new transients, studying the long term evolution 
of persistent sources, monitoring the frequent outbursts of sources in the hard X-rays.
Most of these sources are X-ray binaries, both in low or high mass systems, 
with a black hole or neutron star as compact object.

So far, during the CP eight new transients have been discovered by the CdTe 
detector layer (ISGRI) of IBIS (\cite{lebrun03}).
These sources show different features: quick variability, weak emission 
and soft spectrum as for IGR J16358--4726 (\cite{rev_1635}),
strong flux and hard spectrum in the IGR J17464-3213 case (\cite{rev_1746}).
Usually, after the discovery of a new source, follow up observations (i.e. by XMM-Newton, Chandra, RXTE) 
and searches in the data archives are performed.
For example, IGR J16320-4751 (discovered by INTEGRAL during an open time observation) 
has been observed by XMM-Newton (\cite{rodri03_1}) and it has been consistently detected 
in a reanalysis of the data collected in 1996--2002 with the BeppoSAX-WFC (\cite{zand03}).

During the Performance Verification (PV) phase, the first GPS has been performed in the Cygnus region:
Cyg X--1, Cyg X--2 and Cyg X--3 were expected to be detected and localized.
These three sources have different natures: Cyg X--1 is a black hole candidate in a high mass system, characterized 
by state transitions, and during our observation was in its so-called hard state (\cite{zdzi02} and references therein);
Cyg X--2 is a bright low mass X-ray binary (LMXRB) classified as Z source, showing Type-I X-ray bursts (\cite{kuulk95});
Cyg X--3 is a high mass X-ray binary (HMXRB) which with its typical radio jets is one of the dozen galactic binaries classified 
as microquasar (see list in \cite{diste02}), even though the black hole or neutron star 
nature is still unknown (\cite{hanni03}).
\begin{figure}[!t]
\begin{center}
\epsfig{file=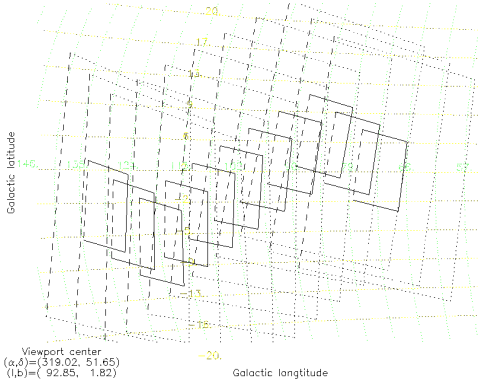,width=6.3cm}
\end{center}
\caption{Scheme of the GPS performed during revolution 26 starting from $l=73^\circ$.} 
\label{fig:map}
\end{figure}

\section{Observations and Data Analysis}

During revolution (rev.) 25 on December $28^{th}$ 2003 and rev. 26 on December $30^{th}$, two scans of the 
Galactic Plane were performed in the Cygnus region with a total of 21 pointings: 11 in rev. 25 and 10 in rev. 26.
    
The difference between the two GPS is that the first scan started from $l=128^\circ$ going 
on for roughly 57$^\circ$ towards the Galactic Centre, 
whereas the second one has been performed in the opposite direction (Fig. \ref{fig:map}). 
The continuous squares in Fig. \ref{fig:map} 
describe the limits of 9$^\circ\times$9$^\circ$ IBIS fully coded field of view (FCFOV); 
the dashed ones define the total instrument field of view: FCFOV plus partially coded field of view (PCFOV).
Details of the first and last scan of each revolution are given in Tab. \ref{tab:details}.
\begin{table}[!t]
\caption{Log of the starting ($sp$) and ending ($ep$) pointings of the two GPSs in revolutions 25 and 26.}
\begin{tabular}{c c c c}
Exposure    &  Duration      & \multicolumn{2}{c}{Pointing coordinates (J2000)}\\
number          &  ({\rm s})    &   {\rm R.A.}       & {\rm Dec}                   \\
\hline
00250026 ($sp$)          & 2263       & $01^h$ $41^m$ $00.00^s$ & $+64^\circ$ $28'$ $58.8''$\\
00250036 ($ep$)   & 2200       & $19^h$ $45^m$ $00.00^s$ & $+37^\circ$ $22'$ $01.2''$\\
00260002 ($sp$)   & 4088       & $20^h$ $07^m$ $00.00^s$ & $+36^\circ$ $22'$ $01.2''$\\
00260011 ($ep$)   & 3781       & $01^h$ $00^m$ $00.00^s$ & $+58^\circ$ $34'$ $01.2''$\\

\hline
\end{tabular}
\label{tab:details}
\end{table}
We analysed consolidated data re-processed by the INTEGRAL Science Data Centre (ISDC) system (\cite{courvo03}),
and we used the IBIS analysis software (\cite{gold03}) delivered by the ISDC to the wide scientific community
on May $21^{th}$ 2003 within the total INTEGRAL Off-line Scientific Analysis software (OSA 1.1).
For each exposure, we have extracted images in three different energy ranges:
20--40 {\rm keV}, 40--80 {\rm keV}, 80--160 {\rm keV}.
In the software release there is a complete X-ray and gamma-ray sources catalogue prepared by \cite{ebi03},
which is used as OSA input in order to distinguish known and new sources.
No new sources have been found in the Cygnus region during the observations presented in this paper.

Furthermore, of the three sources which we expected to see, we have extracted for the strongest, Cyg X--1,
a spectrum for the first pointing of rev. 26, when the source was in the FCFOV. 
We compared its count rate spectrum to that obtained for the Crab in an on-axis observation (rev. 39),
since the spectral responses for fully coded sources are comparable and better known.
In general, the spectral binning is selected giving the response matrix (rmf) as input. For our analysis, the IBIS/ISGRI 
original 2048 linear elements of the rmf have been rebinned to 64 non-linear energy channels.

\section{Results}

\begin{figure}[!t]
\begin {center}
\epsfig{file=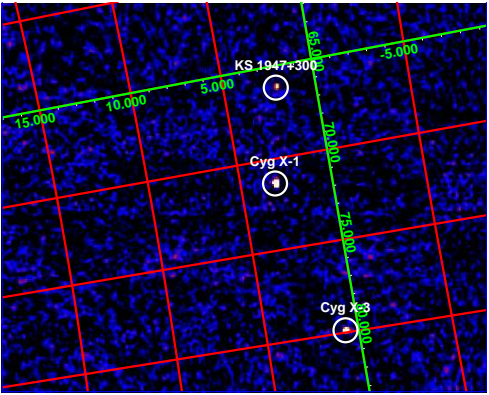,width=6.5 cm}
\caption{Image in galactic coordinates of the last pointing of revolution 25 in the 20--40 {\rm keV} energy band. 
The detection significance of KS 1947+300 is $11\sigma$.} 
\label{fig:1947}
\end {center}
\end{figure}

During the observations, the three expected sources were immediately detected in the preliminary near real time analysis 
(\cite{courvo03}).
Using the last delivered software and performing a deeper analysis, also the transient X-ray pulsar SAX J2103.5+4545 has been 
detected in rev. 25 as reported by \cite{luto03} and was still present in rev. 26.
Moreover, we found in two pointings a statistically significant signal corresponding to the position of the KS 1947+300,
a HMXRB in Be system; 
in Fig.\ref{fig:1947} the 20--40 {\rm keV} image obtained from exposure 00250036 is shown. 
This transient was also revealed in the 40--80 {\rm keV} and 80--160 {\rm keV} bands.

Informations concerning all source detections for the range 20--40 {\rm keV} are given in Tab. \ref{tab:results}.
Of the available 21 pointings, we had source detections in only 8 exposures. 
In the energy range 40--80 {\rm keV}, the three objects Cyg X--1, Cyg X--3 and KS 1947+300 have been detected;
Cyg X--1 and KS 1947+300 are still visible in 80--160 {\rm keV}.
In Tab. \ref{tab:high} the intensities of the three latter sources are given as measured in pointing 00250036.
We found evident flux variations for these sources.

However, in this first phase of the mission, we must be careful with the interpretation of such variations since these
could partly be caused by  systematic effects which have to be investigated further.

\begin{table}[!b]
\caption{Source detections in the Cygnus region during GPS 25 and 26 for the 20-40 {\rm keV} energy range:
source name, flux, offset between the catalogue position and that found with the present software, 
source inclination angle with respect to the pointing direction.}
\begin{center}
\begin{tabular}{c c c c }
$^\S$Source & Flux         & Offset &       Off-axis angle\\
             &({\rm mCrab})& ({\rm arcmin}) &  ($^{\circ}$) \\

\hline
\hline
\multicolumn{4}{c}{Exposure n$^\circ$ 00250032}\\
\hline
Cyg X--3 & $147\pm 8$ & 3.0 & 14.7  \\
Cyg X--2 & $39\pm 5$ & 3.0  & 11.6\\
\hline
\multicolumn{4}{c}{Exposure n$^{\circ}$ 00250033}\\
\hline
Cyg X--3 & $86\pm 3$ & 1.2 & 9.0 \\
J2103.5+4545 & $27\pm2$ & 2.7 & 1.8 \\

\hline
\multicolumn{4}{c}{Exposure n$^{\circ}$ 00250034}\\
\hline
Cyg X--1  & 392 $\pm4$ & 1.5  &  11.9   \\
Cyg X-3   & 151$\pm2$ & 0.6 &  3.7   \\
\hline
\multicolumn{4}{c}{Exposure n$^{\circ}$ 00250035}\\
\hline
Cyg X--1  & 410$\pm3$ & 0.6  &  6.4   \\
Cyg X--3 & 175$\pm3$ & 1.1 &   4.2  \\
\hline
\multicolumn{4}{c}{Exposure n$^{\circ}$ 00250036}\\
\hline
Cyg X--1  & 410$\pm3$ & 0.4  &  3.4   \\
Cyg X--3 & 157$\pm4$ & 1.9   &   9.9  \\
KS 1947+300 & 65$\pm4$ & 1.3 &   7.2 \\
\hline
\multicolumn{4}{c}{Exposure n$^{\circ}$ 00260002}\\
\hline
Cyg X--1  & 390$\pm2$ & 0.6  &  2.1   \\
Cyg X--3 & 150$\pm2$ & 1.7  &   6.7  \\
KS 1947+300 & 55$\pm3$ & 1.4 &   7.2 \\
\hline
\multicolumn{4}{c}{Exposure n$^{\circ}$ 00260003}\\
\hline
Cyg X--1  & 386$\pm3$ & 1.2  &  7.6   \\
Cyg X--3 & 168$\pm2$ & 0.7  &   3.8  \\
\hline
\multicolumn{4}{c}{Exposure n$^{\circ}$ 00260005}\\
\hline
Cyg X--3 & 132$\pm3$ & 1.2 &   11.0  \\
J2103.5+4545 & $22\pm2$ & 2.7 & 1.8 \\
\hline
\hline
\end{tabular}\\
\end {center}
\small{$^\S$ If present in the field of view, all these sources have been detected at least in the lowest energy range.}
\label{tab:results}
\end{table}
Using a Crab calibration observations of 100 {\rm ks} duration, we verified that a spectral fit 
with the known Crab spectrum gives an acceptable $\chi^2$ with only systematic deviations at the $10\%$ level for energies
below 70--100 {\rm keV}.
    
For Cyg X--1 and Cyg X--3 the offset with respect to SIMBAD position vs energy has been evaluated  
in the exposures 00250036 and 00250034 respectively, those are when the two sources are fully coded.
As expected, the point source location accuracy (PSLA) improves for stronger sources (Tab. \ref{tab:psla}).

In order to give a first indication of the IBIS/ISGRI spectral capabilities, 
the Cyg X--1 count rate normalized to the Crab one is plotted in Fig. \ref{fig:spec}.
In spite of the short integration time, Cyg X--1 is detected above 200 {\rm keV} 
(for a complete Cyg X--1 spectral analysis see \cite{laurent03} and \cite{pott03}). 
 
\begin{table}[!t]
\caption{Fluxes of the sources in the exposure number 00250036.}
\begin{center}
\begin{tabular}{c c c c}

Energy range    &  Cyg X--1 & Cyg X--3  & KS 1947+300 \\
 ({\rm keV})         &     ({\rm mCrab})     &   ({\rm mCrab})&     ({\rm mCrab})   \\
\hline
\hline
20--40 &         410$\pm3$    &    157$\pm4$  &    65$\pm4$ \\ 
40--80 &          405$\pm4$   &     64$\pm3$ &       59$\pm4$\\
80--160 &    165$\pm3$  &       --   &          37$\pm4$  \\
\hline
\hline
\end{tabular}
\end {center}
\label{tab:high}
\end{table}
\begin{table}[!h]
\caption{Cyg X--1 and Cyg X--3 in the FCFOV: offset with respect to SIMBAD catalogue position calculated for each energy range.}
\begin {center}
\begin{tabular}{c c c}

Energy range    &  Cyg X--1 offset &    Cyg X--3 offset\\
 ({\rm keV})         &       ({\rm arcmin})     &   ({\rm arcmin}) \\
\hline
\hline
20--40 &     0.4    &     0.6  \\
40--80 &     0.6   &      1.1  \\
80--160 &    0.7   &       --   \\
\hline
\hline
\end{tabular}
\end {center}
\label{tab:psla}
\end{table}

\section{Discussion}
\begin{figure*}[!t]
\centering
\epsfig{file=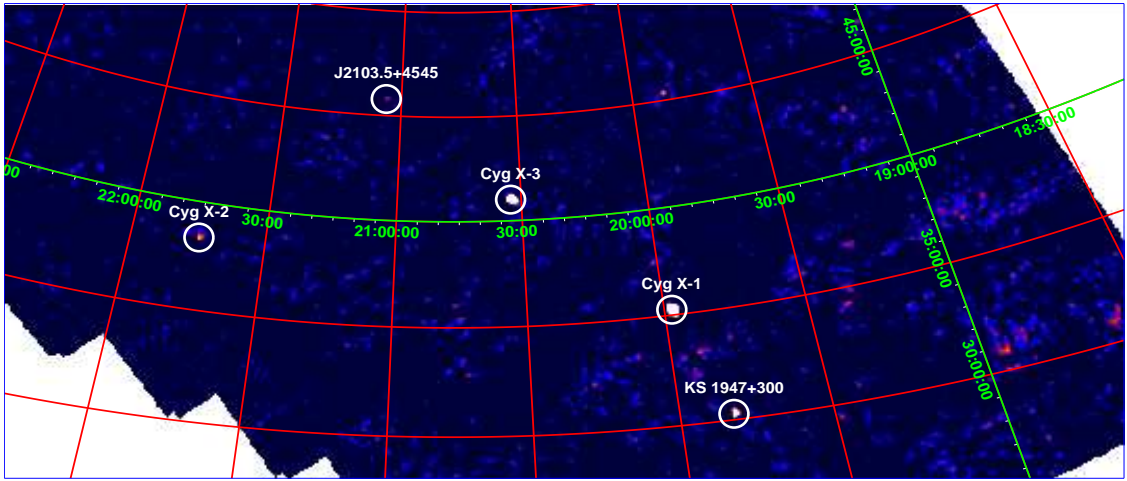,width=15cm, height=6.4cm}
\centering
\caption[]{Mosaic of 20--40 {\rm keV} images of all exposures in Tab. 2.}
\label{fig:mosa}
\end{figure*}

We have demonstrated (see Tab. \ref{tab:results}) that we can give a lower limit
for the IBIS/ISGRI source detection sensitivity ($\sim$5$\sigma$ in 2200{\rm s}) of $\sim$20 {\rm mCrab} in the 
20--40 {\rm keV} band, in good agreement with \cite{rodri03_2}.
Collecting a larger amount of GPS data we expect to detect more transients and new sources.
For example, SAX J2103.5+4545 that is quite imperceptible in one exposure image, becomes visible
on the summed mosaic map (Fig. \ref{fig:mosa}).
\begin{figure}[htb]
\epsfig{file=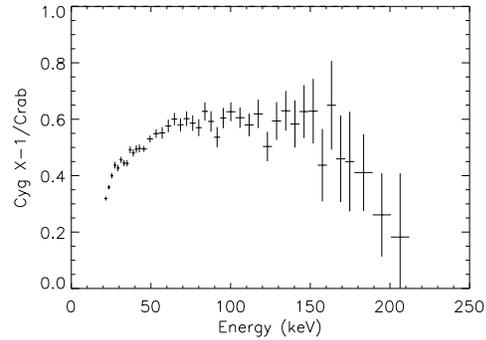,width=7cm}
\caption{Cyg X--1 count rate spectrum (exposure 00260002) normalized to that of the Crab.
The good time interval amounts to roughly 3700 seconds.} 
\label{fig:spec}
\end{figure}
In the short exposures of the GPS, the point source location accuracy (PSLA) 
of faint sources in the FCFOV is 2.7$^{\prime}$ based on the SAX J2103.5+4545 detection.
For a strong source like Cyg X--1, we find an offset of 0.4$^{\prime}$.
As expected, when the source is far away from the pointing direction the PSLA is worse;
in fact Cyg X--1 at an off-axis angle of 11.6$^\circ$  was measured with an offset of 3.0$^{\prime}$.

The statistical errors of the PSLA are significantly larger than the theoretical values and clearly 
dominated by systematics (\cite{gros03}),
mainly due to the misalignment rotation matrix computed at ISDC (\cite{walter03}) in order to correct 
the offset between the telescope axis and the axis of the star sensors.

\indent Finally, during the GPSs of revolutions 25 and 26, IBIS/ISGRI detected 
Cyg X--1 at a flux level of roughly 400 {\rm mCrab} for energies 20--40 {\rm keV}.
The flux was variable, as also reported in \cite{bazzano03}.   
Furthermore from Fig. \ref{fig:spec} the black hole candidate exhibits a harder spectrum than the  
Crab below 50 {\rm keV} and softer above roughly 150 {\rm keV}, which is a confirmation of the  
known spectral shape of Cyg X--1 in its hard state.

\begin{acknowledgements}
This work has been partially supported by the Italian Space Agency (ASI).\\
MDS thanks Aleksandra Gros (CEA--Saclay) for the software support during the data analysis.\\
MDS, FC, LF  thank the INTEGRAL Science Data Centre for the hospitality during some part of this work.\\
JR acknowledges financial support from the French Spatial Agency (CNES).
 \end{acknowledgements}

{}

\end{document}